# On Novel Peer Review System for Academic Journals: Experimental Study Based on Social Computing

Li Liu[1], Qian Wang[2], Zong-Yuan Tan[3], Ning Cai*[1]

[1] School of Artificial Intelligence, Beijing University of Posts and Telecommunications, Beijing, China
E-mail: caining91@tsinghua.org.cn

[2] Library, Beijing University of Posts and Telecommunications, Beijing, China

[3] School of Computer Science and Technology, Donghua University, Shanghai, China

**Abstract:** For improving the performance and effectiveness of peer review, a novel review system is proposed, based on analysis of peer review process for academic journals under a parallel model built via Monte Carlo method. The model can simulate the review, application and acceptance activities of the review systems, in a distributed manner. According to simulation experiments on two distinct review systems respectively, significant advantages manifest for the novel one.

## 1 INTRODUCTION

Peer review is regarded as a cornerstone of scientific publication, and has long been employed by journals to filter submissions with relatively high standard since it ensures that the value of academic output is subjected to scrutiny and critical assessment by scholars [1]. However, the sustainability of peer review has also been debated between scholars [2], for instance, the overall performance of peer review could possibly become worse with the quantity of submissions increasing substantially [3], the biases in peer review can lead to a circumstance where it is difficult to obtain funding or publish innovative or controversial research for some researchers [4], and the peer review systems could be manipulated by some dishonest researchers to publish inferior papers [5].

Prechelt *et al.* [6] conducted a survey and indicated that peer review was deemed somewhat negative by some respondents and most of them appeared more willing to try a alternative peer review regime. Efforts have been paid to seek promising alternatives of peer review or augment certain interventions for improving scientific publication. Faggion [7] pointed out that open peer review is more transparent and might be the best choice to replace the regular review process. Herron [8] indicated that the accuracy of post-publication reader review could surpass traditional single-blind peer review if the quantity of readers is sufficient. Kovanis and Birukou *et al.* [9-10] evaluated and compared certain alternative systems of peer review, e.g., portable and cascade peer review, with efforts also aiming at

Corresponding Author: Ning Cai (caining91@tsinghua.org.cn)

improving academic publication process. In the meantime, for better fairness of review, double-blind peer review has been conducted by *Nature* and its sister journals, since it is regarded as being more impartial in general [11].

The basis of peer review is the review activities by experts. The effectiveness of review process is restricted by the boundness of overall resources available. For instance, the expenditure of peer review in higher education is above one hundred million pounds each year in Britain [12]. Furthermore, peer review is rather time-consuming, which inevitably leads to delay in the scientific publication process. Some journals may take over a year to reach a final decision. Limited resources for peer review can hardly accommodate the exponential growth of manuscripts [13]. Review work would become increasingly challenging, with threat exerted to the efficiency of the peer review system and the sustainability of scientific publication.

The current regular review system has been mostly popular hitherto. However, the final acceptance of each manuscript by any journal somewhat bears randomness. Namely, it is quite common that a manuscript has already been rejected by different journals after repeated submissions, however there still exists one journal that ultimately accepts it for publication. As a reviewer, one sometimes may regrettably find a manuscript once rejected by him but was published in another journal, even with a higher impact factor. Also, a scholar might successively receive invitations from different journals to repetitively review a similar manuscript. Besides, as the amount of submissions continues to increase, it is possible that the overall quality of academic publications would deteriorate instead [3]. One primary reason for the above matters should lie in a paradox between the limitation of review resources and the demand of the increasing scale of submissions.

Considering such a situation, a novel review system is proposed for the sake of improving the performance and effectiveness of peer review, which is in some sense analogous to the idea of open review [14-15]. As the primary approach, a simple model is developed to simulate the creation and submission of manuscripts, application, review, and approval activities of the review systems. Simulation results reveal that, compared against the regular review system, the novel one has prominent advantages. 1) The reviewers earn more time to conduct their review carefully, and thereby the review scores are relatively more objective and precise. 2) The average rejection times before publication are observably reduced. 3) The overall standard of academic journals is not only higher but also steadier.

The research is underlain by the methodology of parallel analysis-based social computing [16-17]. Typical parallel social systems are agent-based cyber systems. Partial features of these systems could be known analytically by virtue of existing and new theories in systems and control, e.g., swarm stability [18-19] and swarm controllability [20-22]. Besides, Monte Carlo experimental models are commonly applied to mine out the potential laws of social phenomena [23-24]. The goal of analysis based on experimental systems is essentially not for thoroughly and quantitatively duplicating and anticipating the activities in actual society, whereas it should still be rather effective to qualitatively testify issues and clarify debates.

There are numerous studies in the literature aiming to investigate and improve peer review through different approaches, including social experiments, especially agent-based. To evaluate the function of peer review in scientific publication, Kovanis *et al.* [25] utilized agent-based modeling (ABM) method with empirical data from journals. Based on [25], some novel alternative schemes [9] of peer review were assessed, for the sake of improving scientific publication. Somewhat analogously, Allesina [26] endeavored to study the peer review systems in a quantitative manner. Thurner *et al.* [27] exploited an agent-based model to research how correct, random, and rational reviewers in the peer review affect overall scientific publication quality. The efficiency of peer review system was assessed by Righi *et al.* [28], for filtering better scientific results. Grimalda *et al.* [29-30] addressed many aspects that affecting the relation between peer review and scientific publication, through ABM built on the Belief-Desire-Intention platform. Squazzoni *et al.* [31-33] developed several simulation scenarios, concerning the consequence of several kinds of interactions between authors and reviewers in the peer review process. Etkin [34] proposed a metric to measure the performance of peer review. To examine the efficiency of peer review, Mrowinski *et al.* [35] created a directed weighted network to represent editorial workflows. Hak *et al.* [36] advocated financial rewards to individuals who participate in review activities frequently, and also proposed an R-index to track reviewer contributions. Based on Monte Carlo experiments, Tan *et al.* [3] revealed the possibility of potential negative effect growing out of excessive submissions. Empirically, Sikdar *et al.* [37] studied three separate entities of the peer-review process (manuscript, author, and reviewer) and discovered several characteristics about predicting the long-term effect of papers. Lendnk [38] suggested to employ gamification techniques and public reviews to match reviewers with the aid of a global reviewer database introduced. Emile [39] proposed an interactive system that allows anonymous communication between scholars and reviewers to improve the efficiency of peer review. Based on Deep Learning techniques, Ghosal *et al.* [40] endeavored to use AI to assist peer review.

Compared with former studies, the novelty, contribution, and feature of the current can be mainly summarized threefold. 1) A novel peer review system is proposed. 2) The advantage of the novel review system is demonstrated explicitly, from several aspects. 3) The effectiveness and applicability of the parallel model originally developed in [3] is further validated through a practical instance.

The organization of the remaining part of this paper is as follows. The framework of the model is described thoroughly in Sec. 2. The comparison analysis regarding the effectiveness of peer review between two systems is conducted in Sec. 3. Additional experiments and sensitivity analysis are presented in Sec. 4 and Sec. 5, respectively. Finally, conclusions are drawn in Sec. 6.

## 2   MODEL FORMULATION

The intermediate platform is the key of the novel system, which is some third-party institution independent of the journals, playing a very significant role in the review process. At the first step, it is the platform instead of the journals who receives the manuscript submissions, directly. Then the platform assigns the submissions to the professional reviewers for review and later release the review scores gathered. After the authors receive the review scores, they will choose several candidate journals to apply for publication, according to the scores known. Finally, the journals select those manuscripts with relatively higher quality and accept them for publication.

For simulating and analyzing such a novel process of peer review, a parallel model is developed via inheriting and extending the model in [3].

This model is discrete-timed, with each iteration representing a round of review and publishing in real-time, i.e., a new issue for all journals. For better understanding, the procedure is divided into several phases.

The first phase is the initialization, which involves setting some basic information and parameter values, namely:

1) The number of journals.

2) The newly increased number of submissions per round, $n$.

3) The number of final publications per round.

4) The total number of reviewers.

5) The quality indicator $\eta$. Each manuscript is randomly allocated a value $\eta \in [0,10]$ indicating its intrinsic quality.

6) The quartile thresholds $\theta_1$, $\theta_2$, $\theta_3$, which represent the lowest level of average quality of journals in quartiles $Q_1$, $Q_2$, $Q_3$, respectively.

7) Certain other relevant parameter values.

The second phase is the setting of journal rankings. Resembling [3], the journals in the model are classified into quartiles, with the first quartile covering the top 10%, the second quartile 10%-25%, the third quartile 25%-50%, and the fourth quartile containing the remainder. In the initial round, a threshold is artificially prescribed for each quartile. After the initial round, these thresholds are iteratively updated by the corresponding data generated in each issue.

The third phase is the creation of the manuscripts. A quantity is said to follow the power-law distribution when the frequency of occurrence of a particular value of that quantity varies inversely as a power of that value. Power-law distribution is widely common in various fields [41], especially for social phenomena. For instance, Barabási *et al*. [42] pointed out that many complex systems such as world-wide web and genetic networks follow power-law distribution. According to Redner [43], the relationship between papers and citation rank is also compatible with power-law distribution. Analogously, we hypothesize the quality of manuscripts follows power-law distribution with exponential cutoff, as illustrated in Eq. (1) and Fig. 1 (a).

$$f(\eta) = a\eta^{-b}e^{-c\eta} \quad (1)$$

where $\eta$ denotes the quality of manuscripts and $f(\eta)$ its probability density function; *a*, *b*, and *c* are parametric constants that determine the shape of curve. To generate a random number with given probability density function $f(\eta)$, one needs to further obtain its probability distribution function $F(\eta)$, with $F(\eta) = \int_0^\eta ax^{-b}e^{-cx}dx$. Here, Monte Carlo method is employed for numerical integration. Fig. 1 (b) shows the resulted probability distribution curve. For each manuscript $i$ ($i = 1, 2, ..., n$), its quality can be generated through mapping an evenly random seed $p \in [0,1]$ on the probability distribution curve. The ultimate distribution of manuscripts is manifested in Fig. 1 (c).

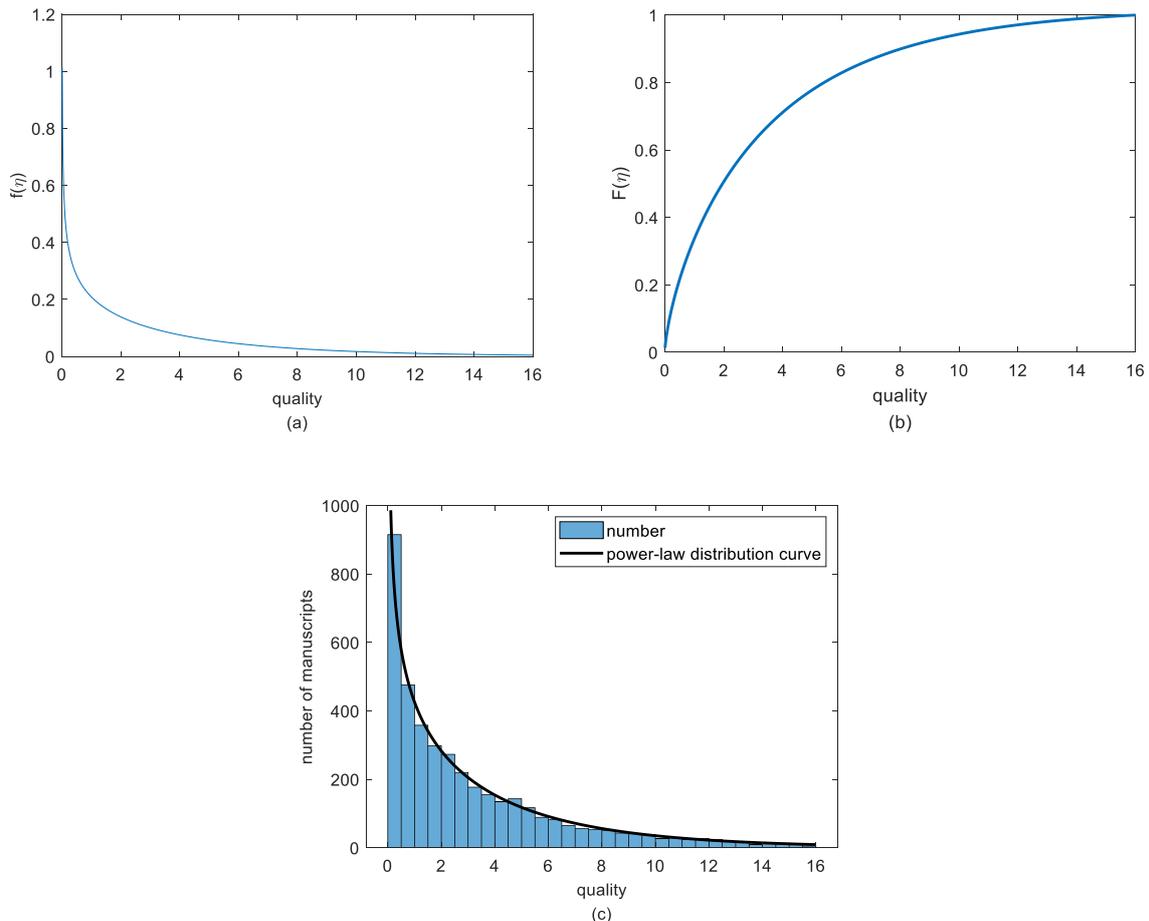

Fig. 1 (a) Probability density curve of power-law distribution with exponential cutoff (a=0.255, b=0.3, and c=0.2), frequency of occurrence decreases with increasing quality; (b) Probability distribution curve; (c) 4000 samples randomly generated by Monte Carlo method under power-law distribution.

The fourth phase is the creation of the intermediate platform, whose main function is to assign the manuscripts for peer review and release the review scores.

The fifth is the peer review process, which is the pivotal phase. Reviewers evaluate the manuscripts through a score system. In the current novel system model, six reviewers are randomly chosen for each manuscript, by the intermediate platform. For revised manuscripts, three reviewers are needed for each manuscript with previous review score being referred to. After a manuscript is assigned for review, the number of manuscripts being simultaneously reviewed by a reviewer correspondingly increases 1. In sharp comparison, only two reviewers are invited to review each manuscript in the regular peer review system, by the journal it is submitted to. Assume that the accuracy of a review score is affected by the number of manuscripts allocated to the reviewer. This is because that, a review is patient and meticulous if only the reviewer has enough time to do this. Accordingly, a relatively objective and precise score can be given. Considering this fact, the score is calculated as follows:

$$\hat{\eta}_i^{(j)} = f(\eta_i) = \eta_i \cdot (1+\delta^{(j)}) \quad (i=1,2,...,n; j=1,2,...,6) \tag{2}$$

where $\hat{\eta}_i^{(j)}$ is an estimator denoting the review score by reviewer $j$ to manuscript $i$; $\eta_i$ denotes the intrinsic quality of manuscript $i$; $\delta^{(j)}$ denotes the multiplicative review noise imposed to reviewer $j$, reflecting the idiosyncrasy of individuals, which is a Gaussian random number with an expectation 0 and a variance $\beta \cdot (n_R^{(j)})^\gamma$, where $n_R^{(j)}$ denotes the number of manuscripts being simultaneously assigned to reviewer $j$, and $\beta, \gamma \in R^+$ are parametric constants that together determine the aggregation or dispersion of distribution. One knows that the variance is positively related to the number of simultaneous reviews, and intuitively speaking, it indicates a measure to the deviation of a reviewer's judgement.

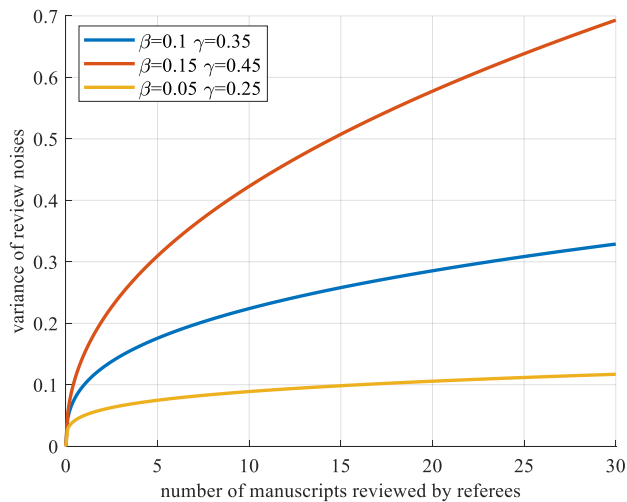

Fig. 2 Variance of review noises varies with the number of manuscripts reviewed by the reviewer

The sixth phase is application, which is analogous to submission in some sense but different. Authors apply for acceptance by the target journals based on their review scores already achieved. Suppose that each manuscript can simultaneously apply for evaluation by two journals lying in different quartiles. Let us consider the potential behavior of a rational author. Initially, he may choose a journal right in the quartile in which the review score falls, meanwhile choose a higher quartile to try his fortune. However, if a manuscript has been rejected more than one or two times, the author would tend to lower his expectations for boosting the chance of success.

The seventh phase is acceptance. In the current model, the journals in higher quartiles have higher priority in selecting. The applications are evaluated and chosen by the journals according to the principle of preferential admission. Journals sorted the applications by the review scores, then a prescribed quantity of them with relatively higher scores are selected. If an application has already been accepted by another journal, this application will be canceled by the current journal.

Finally, the average quality of journals per issue is calculated according to the publication record. Those rejected manuscripts can be revised and resubmitted to other journals in the next round. The intrinsic quality of revised manuscripts will be updated as follows:

$$\eta_i = \eta_i + y_i \tag{3}$$

where $\eta_i$ denotes the intrinsic quality of manuscript $i$; $y_i \sim N(\mu \cdot \alpha^k, \sigma^2)$ is normal distribution with mean $\mu \cdot \alpha^k$ and standard deviation $\sigma$; $k$ is the times that manuscript $i$ have been revised and $0 < \alpha < 1$ is the parameter determining the attenuation trend of mean with $k$.

## 3   COMPARISON ANALYSIS FOR TWO DIFFERENT REVIEW SYSTEMS

In comparison with the regular review system model [3], the current review system has explicit advantages. To verify this, relevant simulations are conducted, demonstrated and analyzed in this section, addressing several major aspects.

Firstly, let us observe and compare the average number of reviews held by one reviewer, as shown in Fig. 3 (a). It can be found that the average number of reviews per reviewer in the novel system is generally lower than the regular system, that is because novel system can effectually reduce the accumulation of low-quality manuscripts which causes reviewers overburdened in regular system and real life, like Fig. 3 (b). Besides, one can see that, as the scale of submissions and issue keeps on increasing persistently, such a difference in the review number is growing more and more intensive. This indicates that the novel system can effectively reduce the working burden of reviewers, and the merit should be more conspicuous with time and high scale of submissions. Thereby, the reviewers have more time to review and understand each manuscript.

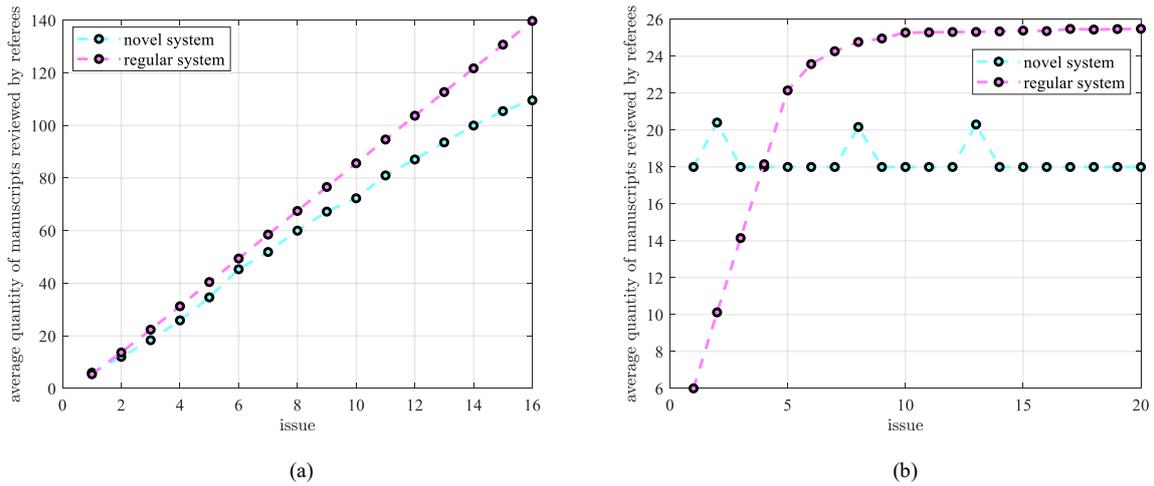

(a) (b)

Fig. 3 Change of average quantity of reviews per reviewer. (a) with *n* (issue=20); (b) with issue (n=3000).

Secondly, let us consider the common but somewhat annoying fact: acceptance is not always right the first time. Usually, a publication has to experience several times of submissions and rejections before receiving an ultimate acceptance. The average times of rejections per publication is also a meaningful indicator. Fig. 4 illustrate the variation of this indicator in two peer review systems.

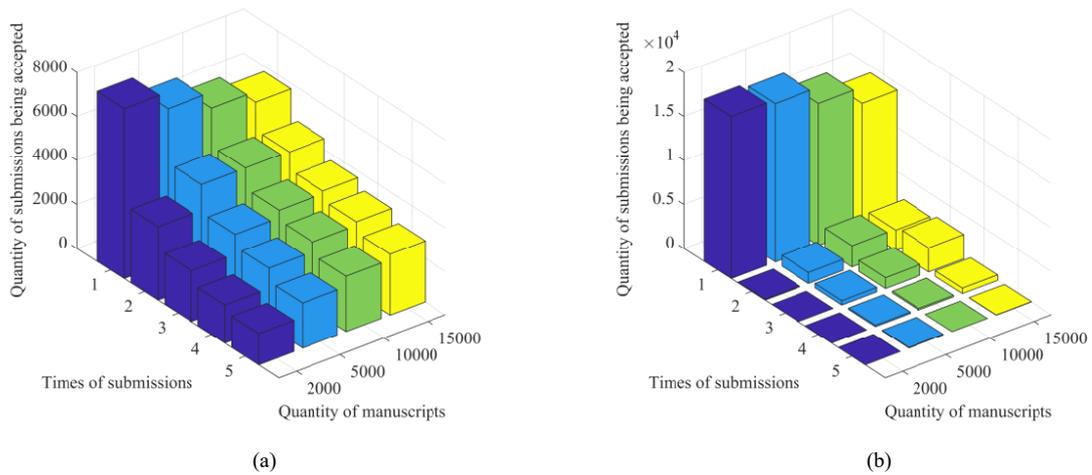

(a) (b)

Fig. 4 (a) Variation of quantity of submissions accepted and times of submissions with different *n* (regular system); (b) Variation of quantity of submissions accepted and times of submissions with different *n* (novel system).

By observing Figs. 4, one can clearly sense that in both systems, the number of acceptances being right the first time descends as the scale of submissions keeps on increasing, however, the change of the scale of submissions has little effect on novel system, but has great effect on regular system. In the regular system, acceptance occurs mostly by the fourth and fifth round of submissions if with sufficient submissions; while in comparison, most of the publications are accepted by the first round in the novel system. The number of acceptances being right the first time is also greater in the novel system. Moreover, in general, the count of rejections before acceptance never exceeds three in the novel system, whereas it might reach up to five in the regular system. In sum, the novel system is more advantageous than the regular one, in regard of the average number of rejections per publication as an indicator.

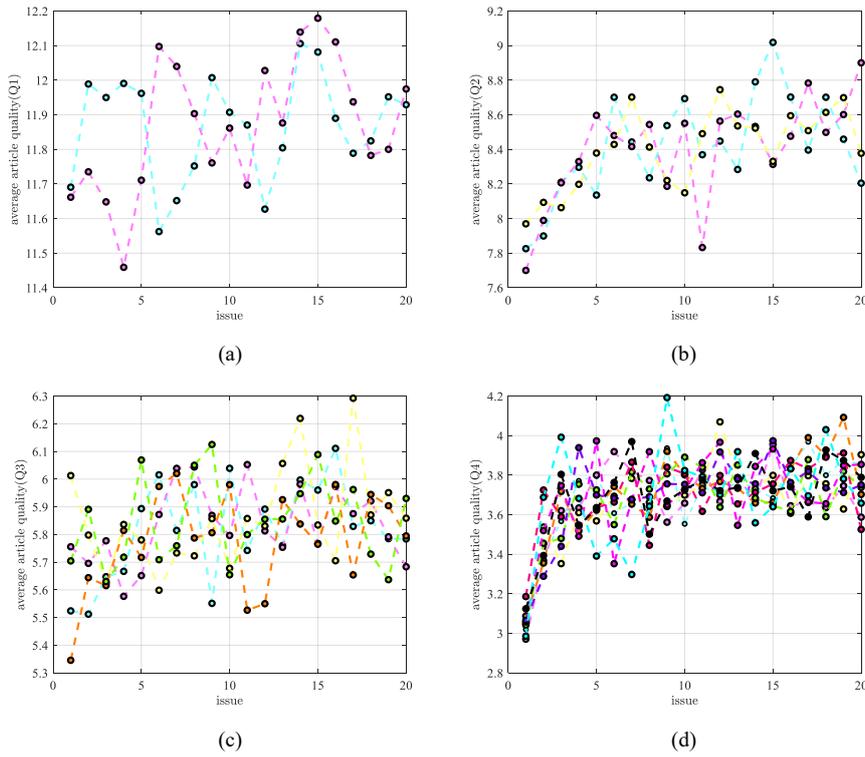

Fig. 5 Variation of average quality of journals in quartiles with issue, where *n*=3000 (regular system). (a) Q1; (b) Q2; (c) Q3; (d) Q4.

Thirdly, the average standard of journals will be influenced by the burden of reviewers and rationality of the accepting manuscripts, therefore the average standard of journals is a meaningful indicator that can distinguish the advantage between peer review systems. In order to analyze, the trends of average journal quality in different quartiles are simulated, as Figs. 5 & 6.

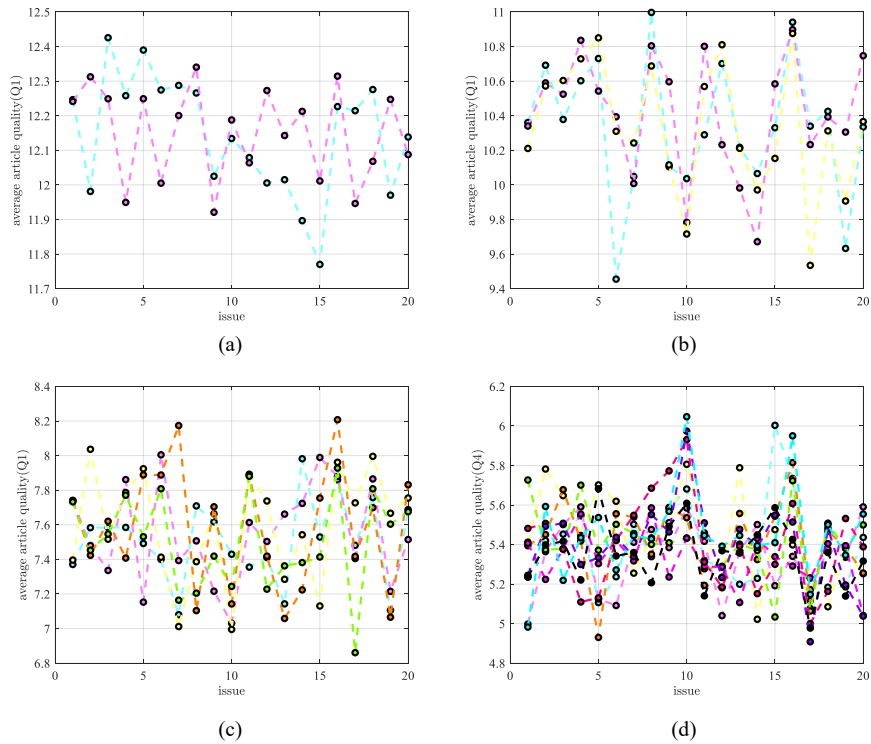

Fig. 6 Variation of average quality of journals in quartiles with issue, where *n*=3000 (novel system). (a) Q1; (b) Q2; (c) Q3; (d) Q4.

In sharp contrast, for the novel peer review system, in Q1 and Q2 the journal average standards have narrower range of variation and are stabler (although with some ups and downs) compared with the regular system, while the journal average standards in Q3 and Q4 also easily enter a steady state in both systems. Compared with the regular system, for the novel system in Q1, Q2 and Q3, the journal average standard is markedly higher and keeps relatively steadier, but in Q4 the journal average standard is similar and sometimes even lower than the regular system.

What accounts for this difference? On one hand, the difference of burden for reviewers causes the difference of stationarity between the two systems. On the other hand, the advantage of intermediate platform is the reason of higher journal average standard in novel system. For the regular system, the fact that some high-quality manuscripts can be found in Q4 may cause such a result that the journal average standard is similar and sometimes even higher than the novel system. This also implies the weakness of the regular system. The intermediate platform can enhance the rationality of the manuscript acceptances, such that papers at all levels are included in appropriate journals, through filtering out low-quality manuscripts and reducing the burden of reviewers. These indicates that the novel system also holds distinct advantage in being robust against the magnitude of submission amount. In other words, the novel system manifests better effectiveness in filtering out low-quality manuscripts, keeping high-quality manuscripts and further guaranteeing the overall standard of journals.

## 4  SIMPLIFIED SUBMISSION RULE

Next, let us simplify the simulation rule for the evolution of journal quality under the novel peer review system. The consideration of quartiles of journals is bypassed, instead, authors submit straightly according to the quality of their manuscripts and the quality of journals. They can choose two journals with the quality nearest to the estimation of their manuscripts, one journal being above their manuscript's quality and another below. In the first issue of experiment, assume that all journals have the same quality and all manuscripts are randomly assigned to journals, afterwards, they will evolve autonomously. Fig. 7 illustrates the evolution of journal quality in 40 issues.

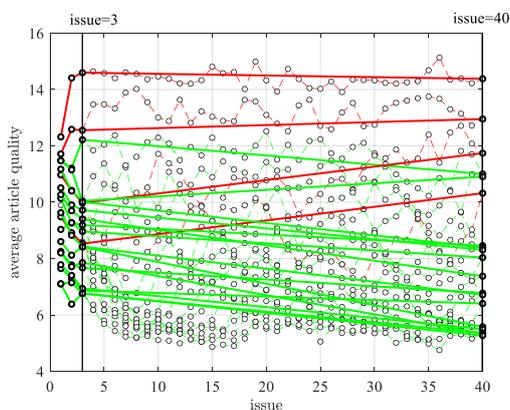

Fig. 7 Variation of average quality of journals with issue, where $n$=10000.

By observing Fig. 7, one can see after the first issue, the journals with initially higher quality will have faster increase in quality and becomes steady earlier, where for visual clarity, the referential trends are marked by thick lines. This indicated that Matthew effect exists in the development process of journals. High quality strengthens the tendency of continuous rising, and vice versa. Then we change *n* to observe the change of journal quality under different manuscript quantity, as illustrated in Fig. 8.

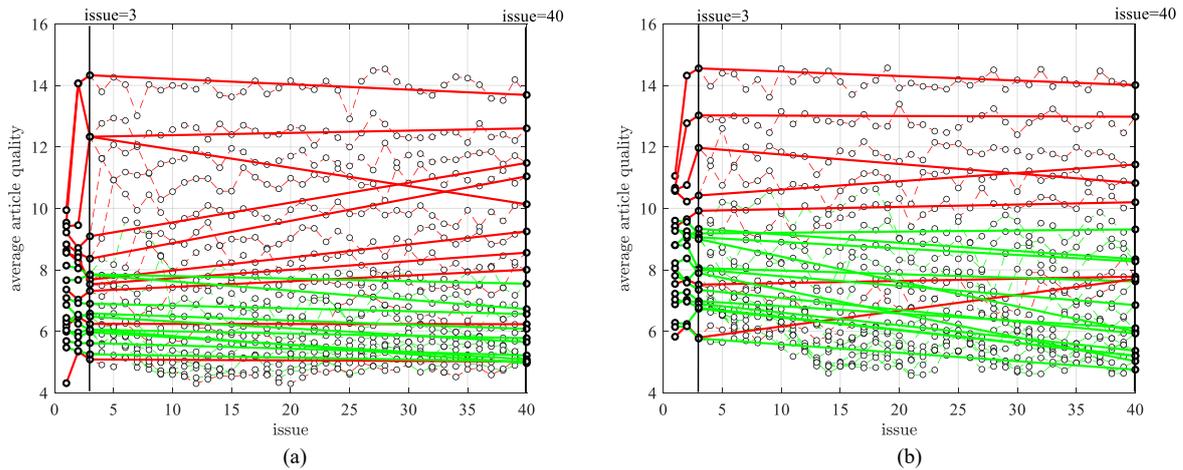

Fig. 8 Variation of average quality of journals with issue, where *n*=5000 (a) and n=7000 (b).

It can be seen from Figs. 7 & 8 that the trends of average quality of journals are very similar, no matter *n* = 5000, *n* = 7000 or *n* = 10000. There are approximately ten percent of journals being of the highest quality, fifty percent being of the lowest, and the remaining above and below average, with the formation roughly analogous to quartiles also.

## 5   SENSITIVITY ANALYSES

We performed sensitivity analyses for the experimental system by selecting Eqs. (1), (2) and (3) and eight kernel parameters. To evaluate how those variables affect the outputs of the novel system, we explored certain range of values for each variable and assessed robustness of the model, considering the following aspects: the burden of reviewers (Average quantity of manuscripts reviewed by each reviewer), average quality of journals in Q1, Q2, Q3 and Q4 and the probability of manuscripts be accepted by the first time. Secondly, we explored various scenarios that incorporated manuscripts of different quality ranges, different reviewer levels, the number of reviews per reviewer, and authors' ability to revise manuscripts. All sensitivity analyses were averaged over 10 simulation runs.

Eq. (1) is employed to generate the quality distribution of manuscripts, where different *a*, *b* and *c* can indicate different quality level of manuscripts.

**Table 1** Sensitivity analyses defined by varying three parameters of Eq. (1)

| Parameter names | Range of variation | Step of variation | Q1 | Q2 | Q3 | Q4 | The burden of reviewers | The probability of manuscripts be accepted by the first time |
|---|---|---|---|---|---|---|---|---|
| $a$ | [7-9] | 0.2 | [11.94-12.31] | [9.32-11.18] | [6.75-8.61] | [4.70-6.00] | [30.2-31.1] | [0.955-0.984] |
| $b$ | [0.2-0.4] | 0.02 | [11.68-12.25] | [9.33-11.08] | [7.31-7.7.] | [5.14-5.56] | [30.2-31.3] | [0.942-0.986] |
| $c$ | [0.1-0.3] | 0.02 | [12.10-12.25] | [10.01-10.50] | [7.04-8.10] | [4.67-6.16] | [30.2-31.6] | [0.929-0.983] |

Range of desired outputs [min–max] from sensitivity analysis. The variation of $c$ is presented as a max to min value, because the highest value of $c$ corresponds to the lowest output results and vice versa.

**Table 2** Sensitivity analyses defined by varying two parameters of Eq. (2)

| Parameter names | Range of variation | Step of variation | Q1 | Q2 | Q3 | Q4 | The burden of reviewers | The probability of manuscripts be accepted by the first time |
|---|---|---|---|---|---|---|---|---|
| $\beta$ | [0.02-0.20] | 0.02 | [11.55-12.51] | [9.66-10.27] | [7.36-7.60] | [5.30-5.40] | [31.8-36.1] | [0.900-0.964] |
| $\gamma$ | [0.25-0.45] | 0.02 | [11.91-12.27] | [10.09-10.38] | [7.45-7.58] | [5.31-5.37] | [31.1-35.4] | [0.888-0.975] |

Range of desired outputs [min–max] from sensitivity analysis. The variation of $\beta$ and $\gamma$ is presented as a max to min value, because the highest value of $\beta$ and $\gamma$ corresponds to the lowest output results and vice versa.

Eq. (2) denotes the multiplicative review noise of reviewers, where different $a$ represents different academic level of reviewers.

Eq. (3) with different $\mu, \alpha$ and $\sigma$ reflects author's ability to revise their manuscripts.

**Table 3** Sensitivity analyses defined by varying three parameters of Eq. (3)

| Parameter names | Range of variation | Step of variation | Q1 | Q2 | Q3 | Q4 | The burden of reviewers | The probability of manuscripts be accepted by the first time |
|---|---|---|---|---|---|---|---|---|
| $\mu$ | [0.25-0.75] | 0.05 | [12.07-12.12] | [10.17-10.28] | [7.48-7.57] | [5.27-5.41] | [31.3-35.6] | [0.902-0.965] |
| $\alpha$ | [0.6-0.9] | 0.03 | [12.07-12.14] | [10.19-10.26] | [7.49-7.56] | [5.28-5.38] | [31.9-35.7] | [0.890-0.977] |
| $\sigma$ | [0.3-0.8] | 0.06 | [12.10-12.12] | [10.18-10.32] | [7.46-7.60] | [5.30-5.40] | [31.6-35.3] | [0.894-0.960] |

Range of desired outputs [min–max] from sensitivity analysis. The variation of $\alpha$ is presented as a max to min value, because the highest value of $\alpha$ corresponds to the lowest output results and vice versa.

# 6 CONCLUSION

Peer review is very important as a mechanism for quality control of the academic journals. However, drawbacks exist in the system of peer review as our previous research has revealed. An innovative review system is proposed in the current paper to improve the performance and effectiveness of peer review. The key is lying in adding an intermediate platform and the peer review will be arranged by the platform. Social computing based on Monte Carlo methodology is adopted as the main approach for evaluating and analyzing the activities in two different review systems, such as submission, review, and acceptance. The study shows the following results:

(1) The review reports are more objective and accurate in the novel review system. This can be

~12~

attributed to two primary reasons. First, the burden of reviewers is effectively reduced in the novel system, thus they have more time to carefully conduct the review work. Second, for any manuscript, the number of reviewers is several times higher in the novel system, thus the potential randomness is lowered.

(2) Compared against the regular peer review system, the average number of rejections before acceptance is significantly reduced and the probability of manuscripts be accepted by the first time is observably higher in the novel peer review system. This can be attributed to the advantages of the platform. The existence of the platform has changed the original manner of peer review, thus authors can apply for acceptance by the target journals based on their already achieved review scores. To some extent, the change brought by the platform eliminates the deviation of the author's personal estimation of his manuscript quality.

(3) The Matthew Effect does exist in the novel systems. In the beginning, the journal with high quality will grow fast. In the end, the quantity distribution of journals with different quality appears to follow a certain formation of distribution.

Generally speaking, the novel review system proposed essentially holds considerable advantages and prospects.

**ACKNOWLEDGMENTS**

This work is supported by National Natural Science Foundation (NNSF) of China (Grant 61867005), by BUPT Innovation and Entrepreneurship Support Program (Grant 2021-YC-A269), by Fundamental Research Funds for the Central Universities (Grant 2019RC29), and by the Gansu Provincial First-Class Discipline Program of Northwest Minzu University (Grant 11080305).

**COMPETING INTERESTS**

The authors declare that they have no competing interests regarding the publication of this paper.

**DATA AVAILABILITY**

The data in this paper is generated via simulations.